\documentstyle[11pt]{article} 

\begin{document}

\newcommand{\tiN}{\raisebox{-6.5pt}{$\displaystyle
\stackrel{\displaystyle N}{\sim}$}}
\newcommand{\tiM}{\raisebox{-6.5pt}{$\displaystyle
\stackrel{\displaystyle M}{\sim}$}}
\newcommand{\beq}{\begin{equation}} \newcommand{\eeq}{\end{equation}}

\title{
Canonical quantization of a minisuperspace model for gravity using
self-dual variables\thanks{Work supported by the Graduierten-Programm of the
DFG}}
\author{T. Thiemann
\thanks{Center for Gravitational Physics
and Geometry, Pennsylvania State University, University Park, PA 16802-6300,
U.S.A., e-mail : thiemann@phys.psu.edu}}
\date{}

\maketitle



The present article summarizes the work of the papers \cite{1} dealing with the
quantization of pure gravity and gravity coupled to a 
Maxwell field and a cosmological constant in presence of spherical symmetry.\\
Let us stress the following : the motivation for this project was {\em not}
to quantize a black hole. Rather, we regard the present model as an
interesting testing ground for the quantization of full 3+1 gravity, in
particular by using Ashtekar's self-dual representation.\\
Throughout we assume that the reader is familiar with the Ashtekar-formulation
of gravity (\cite{2}). The conventions are as in \cite{1}.

To reduce gravity and our matter content to spherical symmetry,
we require that the 3-metric and the Maxwell electric ($\epsilon^a$) and 
magnetic 
fields ($\mu^a$)
are Lie annihilated by the generators of the SO(3)
Killing group. 
The result of these Killing-reduction prescriptions is the 
following :\\
Denoting the angular variables by 
$\theta,\phi$, the radial variable by x , and the standard orthonormal basis on
the sphere by $\{n^a\}$ we obtain for the 
gravitational and Maxwell sector respectively ($A_I,\; E^I,\; I=1,2,3$ are 
angle-independent functions of x and t) 
\begin{eqnarray}
(E^x_i,E^\theta_i,E^\phi_i) & = & (E^1 n^x_i \sin(\theta),\frac{\sin(\theta)}
{\sqrt{2}}(E^2 n^\theta_i+E^3 n^\phi_i),\frac{1}{\sqrt{2}}(E^2 n^\phi_i-E^3 
n^\theta_i)),\nonumber\\
(A_x^i,A_\theta^i,A_\phi^i) & = & (A_1 n^x_i,\frac{1}{\sqrt{2}}(A_2 n^\theta_i
+(A_3-\sqrt{2})n^\phi_i),\frac{\sin(\theta)}{\sqrt{2}}(A_2 n^\phi_i-(A_3-
\sqrt{2})n^\theta_i)\\
(\epsilon^x,\epsilon^\theta,\epsilon^\phi)& :=& (\epsilon(x,t),0,0),\;
(\mu^x,\mu^\theta,\mu^\phi) := (\mu(x,t),0,0) .
\end{eqnarray}
The Maxwell
potential is thus given by
$(\omega_x,\omega_\theta,\omega_\phi)=(\omega(x,t),0,0)
+(\Omega_a(x,t,\theta,\phi))$, 
where $\Omega_a$ is a monopole solution with charge $\mu$. 
The cosmological constant will be labelled by the (real) parameter $\lambda$
and by performing a 'duality rotation' we get rid of the magnetic charge. 
\\
The reality conditions tell us that $E^I,\;i(A_I-\Gamma_I),\;p',\;\omega$ are 
real
where $\Gamma_1=-(\mbox{arctan}(E^3/E^2))',\;\Gamma_2=-(E^1)' E^3/E,\;
\Gamma_3=(E^1)' E^2/E$ denote the spherically symmetric components of the
spin-connection
($E:=(E^2)^2+(E^3)^2$).\\
The model has 4 canonical pairs $(\omega,p\; ;\; A_I,E^I)$ and is
subject to the 4 first class constraints (so that the reduced phase space is 
finite
dimensional)
\begin{eqnarray}
^M{\cal G} &=& p'\mbox{ : Maxwell-Gauss constraint},\\
^E{\cal G} &=&(E^1)'+A_2 E^3-A_3 E^2\mbox{ : Einstein Gauss constraint},
\nonumber\\ 
V &=& B^2 E^3-B^3 E^2\mbox{ : Vector constraint},\nonumber\\ 
C &=&(B^2 E^2+B^3 E^3)E^1+\frac{1}{2}((E^2)^2+(E^3)^2)(B^1
+\kappa\frac{p^2}{2 E^1}+\kappa\lambda E^1)
\mbox{ : Scalar constr.}\nonumber
\end{eqnarray}
where we have abbreviated the components of the gravitational magnetic fields 
by $B^I$
($B^1=1/2((A_2)^2+(A_3)^2),\;B^2=(A_3)'+A_1 A_2,\;B^3=-(A_2)'+A_1A_3)$).

For spherically symmetric systems, the topology of the 3 manifold is
necessarily of the form $\Sigma^{(3)}=S^2\times\Sigma$ where $\Sigma$ is the 
1-dimensional manifold (restrictinig ourselves to asymptotically flat 
topologies in which case suitable boundary conditions on the fields are
imposed)) 
$ \Sigma=\Sigma_n \; ,\; \Sigma_n\cong K\cup\;\bigcup_{A=1}^n\Sigma_A \; ,$
i.e. the hypersurface is the union of a compact set K 
and a collection of asymptotic regions (each of which is diffeomorphic
to the positive real line without a compact interval)
with outward orientation and all of them are joined to K. The case of two
ends is physically most interesting.\\ 

We now apply the method of symplectic reduction 
(\cite{7}).
Choosing 'cylinder coordinates'
$ (A_2,A_3)=\sqrt{A}(\cos(\alpha),\sin(\alpha)),\;(E^2,E^3)=
\sqrt{E}(\cos(\beta),\sin(\beta))$ 
it is easy to see that the Gauss-reduced symplectic potential becomes
\beq i\kappa\Theta[\partial_t]=\int_\Sigma dx(\dot{\gamma}\pi_\gamma
+\dot{B^1}\pi_1+\dot{\omega}p(i\kappa)) \; , \eeq
where $\gamma:=A_1+\alpha',\pi_\gamma:=E^1,B^1:=\frac{1}{2}(A-2)\;
\mbox{and}\; \pi_1:=\sqrt{E/A}\cos(\alpha-\beta)$.\\
In the following p will already be taken as a constant. Also we will deal
with an arbitrary cosmological constant for the sake of generality.\\
We take then the following linear combinations of the vector and the
scalar constraint functional
 $E^1 E^2 V+E^3 C\mbox{ and } 
-E^1 E^3 V+E^2 C $
and set these expressions strongly zero. For non-degenerate metrics
($E\not=0$) 
we can now solve for $E^2,\; E^3$ 
and insert this solution into the Gauss constraint. The final solution is
given by  
\beq  [\kappa(-p^2+\lambda(E^1)^2/3)+B^1 E^1]^2=m^2 E^1 \; .
\eeq
The integration constant, m, is real and can be shown to coincide with the 
gravitational mass up to a factor.\\
Equation (5) is an algebraic equation of fourth order in terms of $E^1$ and
therefore very unpractical to handle.\\
The idea is to change the polarization and to chose $B^1$ as a momentum. Then
(5) can be easily solved for $B^1$ and the vector constraint for $\gamma$.
The result for the reduced symplectic potential is given by (modulo a total 
differential)
\beq
\Theta[\partial_t]=  
 \dot{p}\int_\Sigma dx(-i\frac{p\pi_1}{\pi_\gamma}-\omega)
+\dot{m}\int_\Sigma dx(-i/\kappa)\frac{\pi_1}{\sqrt{\pi_\gamma}}
=: \sum_A[\dot{\Phi}_A p_A+\dot{T}_A m_A].
\eeq
It can be shown that $\pi_1$ is imaginary while $\pi_\gamma=E^1$ is real. 
Accordingly, 
T and $\Phi$ are both real 
and resulting reduced phase space can be described as follows : in every
asymptotic end A we have a cotangent bundle over $R^2$.
The treatment for K is similar.\\

One can explicitly check that the variables $m,T,p,\Phi$ commute with all the 
constraints for Lagrange multipliers of comact support, that is, they are
Dirac observables.\\ The reduced action takes the form
$S_{red}=\sum_A[\dot{T}_A m_A+\dot{\Phi}_A p_A-H_{red;A}]\mbox{ where }
H_{red;A}=N_a m_A+U_A p_A$ is the reduced Hamiltonian and we 
have defined
$ N_A(t):=N(x=\partial\Sigma_A,t)\;,\;U_A(t):=U(x=\partial\Sigma_A,t)$ 
where $N:=\det(q)^{1/2}\tiN$ is the lapse function, $G:=\int dx[\Lambda
(^E{\cal G})+U(^M{\cal G})+N^x V+\tiN C]$ being the constraint 
generator. The solution of the equations of motion  
can be written
($\frac{d}{dt}\tau_A=N_A\;\mbox{and}\;\frac{d}{dt}\phi_A=U_A$) 
\beq
m_A(t)  =  \mbox{const.},\; 
T_A(t)  =  \mbox{const.}+\tau_A(t),\; 
p_A(t)  =  \mbox{const.},\;
\Phi_A(t)  =  \mbox{const.}+\phi_A(t) 
\eeq
i.e. the reduced system adopts the form of an integrable system where the
role of the action variables is played by the masses and the charges whereas
their conjugate variables take the role of the angle variables.\\
What now is the interpretation of this second set of conjugate variables ?
The interpretation of m and p follows simply from the fact that they can
be derived from the reduced Hamiltonian, i.e. they are the well-known surface
integrals ADM-energy and Maxwell-charge. However, their conjugate partners
are genuine volume integrals and we are not able to write them as known
surface integrals.
The solution (7) allows for the interpretation that T is the eigentime at 
spatial infinity while
$\Phi$ plays the same role as the variable conjugate to the electric charge
of 1+1 Maxwell theory.

We finally come to the quantization of the system. We follow the group
theoretical quantization scheme (\cite{4}). \\
The phase space for every end is just the cotangent bundle over the 
two-plane so the unique Hilbert-space is the usual one : $L_2(R^2,d^2x)$.
The Schroedinger-
equation in the polarization in which
eigentime and the flux act by multiplication and the mass and the charge by
differentiation becomes unambiguously
\beq i\hbar\frac{\partial}{\partial t}\Psi(t;\{T_A\},\{\Phi_A\})
     =(-i\hbar\sum_{A=1}^n[N_A(t)\frac{\partial}{\partial T_A}
     +U_A(t)\frac{\partial}{\partial \Phi_A}])\Psi(t;\{T_A\},\{\Phi_A\}) \; .
\eeq
It can be solved trivially by separation :
$ \Psi(t;\{m_A\},\{\Phi_A\}):=\prod_{A=1}^n\psi_A(t,m_A,\Phi_A)$ 
and by using the functions $\tau_A,\;\phi_A$ : 
\beq \psi_A(t,T_A,\phi_A)=C_A\exp(k_A\frac{i}{\hbar}[T_A-\tau_A(t)])
                         \times\exp(l_A\frac{i}{\hbar}[\Phi_A-\phi_A(t)])
\eeq
where $C_A$ is a complex number, whereas $k_A,l_A$ must be
real because the spectrum of the momenta, which are self-adjoint, is real.

Final remark :\\
It is interesting to express the observables found in terms of the 
Ashtekar-connection :
Restricting to the case $\lambda=0$ we obtain 
for the integrand of the variables T and
$\Phi+\int_\Sigma dx \omega$ respectively
\beq
-2\frac{A_1+[\arctan(\frac{A_3}{A_2})]'}{(2B^1)^{2-n}}\frac{[p^2\kappa+
\frac{m^2}{B^1}+\sqrt{[p^2\kappa+\frac{m^2}{B^1}]^2-[p^2\kappa]^2})]^{2-n}}
{p^2\kappa+1/2(\frac{m^2}{B^1}+\sqrt{[p^2\kappa+\frac{m^2}{B^1}]^2
-[p^2\kappa]^2})}
\eeq
where $n=1/2$ and $n=1$ respectively and $B^1=1/2((A_2)^2+(A_3)^2-2)$.

\end{document}